\documentclass{emulateapj}

  \usepackage{amscd}
  \usepackage{amsmath}
  \usepackage{amssymb}
  \usepackage[T1]{fontenc}
  \usepackage[varg]{txfonts}
  \usepackage{verbatim}
  \usepackage{array}

  \newcommand{\sn}{SN~Ia}

  \shorttitle{Off-center ignition in SNe~Ia}
  \shortauthors{R{\"o}pke, Woosley and Hillebrandt}

\begin{document}

\title{Off-center ignition in type Ia supernova:\\
I. Initial evolution and implications for delayed detonation}

\author{F. K. R{\"o}pke\altaffilmark{1,2}, S. E. Woosley\altaffilmark{1} and W. Hillebrandt\altaffilmark{2}}
\altaffiltext{1}{Department of Astronomy and Astrophysics, University of
       California, Santa Cruz, CA 95064, U.S.A.}

\altaffiltext{2}{Max-Planck-Institut f\"ur Astrophysik,
              Karl-Schwarzschild-Str. 1,\\ D-85741 Garching, Germany}

\begin{abstract}
The explosion of a carbon-oxygen white dwarf as a Type Ia supernova is
known to be sensitive to the manner in which the burning is
ignited. Studies of the pre-supernova evolution suggest asymmetric,
off-center ignition, and here we explore its consequences in two- and
three-dimensional simulations.  Compared with centrally ignited
models, one-sided ignitions initially burn less and release less
energy. For the distributions of ignition points studied, ignition
within two hemispheres typically leads to the unbinding of the white
dwarf, while ignition within a small fraction of one hemisphere does
not.  We also examine the spreading of the blast over the surface of
the white dwarf that occurs as the first plumes of burning erupt from
the star. In particular, our studies test whether the collision of
strong compressional waves can trigger a detonation on the far side of
the star as has been suggested by \citet{plewa2004a}. The maximum
temperature reached in these collisions is sensitive to how much
burning and expansion has already gone on, and to the dimensionality
of the calculation. Though detonations are sometimes observed in 2D
models, none ever happens in the corresponding 3D
calculations. Collisions between the expansion fronts of multiple
bubbles also seem, in the usual case, unable to ignite a detonation.
``Gravitationally confined detonation'' is therefore not a robust
mechanism for the explosion. Detonation may still be possible in these
models however, either following a pulsation or by spontaneous
detonation if the turbulent energy is high enough.
\end{abstract}



\section{Introduction}
\label{intro_sect}

In the currently favored model for Type Ia supernovae (SNe~Ia), a
carbon-oxygen white dwarf (WD) grows to almost the Chandrasekhar mass,
then explodes due to a thermonuclear instability
\citep[e.g.,][]{hillebrandt2000a}. While the modeling of the explosion
itself has reached a high level of sophistication, with
multi-dimensional studies being routinely carried out by several
groups \citep[e.g.][]{reinecke2002d,gamezo2003a,plewa2004a,roepke2004c,
  roepke2005b, garcia2005a, roepke2006b}, the initial conditions of
this process remain largely unknown. This is unfortunate since the
geometry of the flame ignition has a large effect on the explosion
strength \citep{niemeyer1996a,livne2005a,roepke2006a,schmidt2006a}.

Runaway commences once the WD has accreted sufficient matter from a
binary companion to approach a central density $\sim 3 \times 10^9$ g
cm$^{-3}$ where plasma neutrino losses are exceeded by energy
generation from a highly screened carbon fusion reaction.  The stage
for flame ignition is set by a century of convective carbon burning in
the progenitor WD. It remains unclear, however, if the
first sparks to develop a nearly discontinuous temperature gradient on
their perimeters (the ``flame'') are concentrated in the center of the
star \citep{hoeflich2002a} or spread around by the convective flow in
which they are embedded. Since this convective flow may have a dipole
character, one natural possibility is lopsided ignition displaced
somewhat off-center \citep[e.g.,][]{woosley2004a}.

It is known that, once born, the flame experiences an extended period
of subsonic propagation---a ``deflagration''
\citep{nomoto1976a}. Prompt detonation is
excluded on the grounds that it would produce spectroscopy,
nucleosynthesis, and a light curve very different from observations
\citep[see][for a review]{filippenko1997a}. It would also require a
degree of isothermality in the core that would be very difficult to
achieve \citep{woosley1990a}. The deflagration poses a computational
challenge since the ashes of the burning are buoyant, and that leads to
instabilities and turbulence that can only be followed with any
accuracy in a multi-dimensional calculation. The difficulty is
compounded by the large range of spatial scales---sub-millimeter for the flame
width and Kolmogorov scale to 2,000 km for the WD---and the
high Reynolds number, $Re \sim 10^{14}$.

Calculations of the same WD differ in outcome because of the
assumptions about ignition, various techniques used to treat flame
instabilities, and turbulence. It is not practical to resolve both the
flame and the star, so full-star models, such as the ones presented
here, rely upon an effective flame model and a subgrid scale model for
the turbulence. Qualitatively, the flame advances radially at a speed
given by the flotation of the largest plumes, but the lateral
spreading of each plume and the overall efficiency of the explosion
can vary, depending upon the way turbulence is handled, and on the
dimensionality and resolution of the model.  Results of different
approaches have been published in a variety of studies with the
general conclusion that a pure deflagration can give a viable
explosion \citep{reinecke2002d,gamezo2003a,roepke2005b}, not too
different from what is observed
\citep{travaglio2004a,blinnikov2006a}. It remains controversial,
however, if these models can give light curves as bright as some
observations indicate, or can explain all of the spectroscopic
features \citep{gamezo2003a, kozma2005a}.

Moreover, these successful models all have in common the assumption of
nearly isotropic central ignition. It may be that nature provides only
anisotropic ignition conditions, or it may be that the the
observational constraints on a pure deflagration will ultimately prove
too stringent. In these cases, a transition to detonation may need to
occur \citep{plewa2004a,livne2005a}.  The idea of a delayed detonation
has been around for some time \citep{khokhlov1991a,woosley1994a}, but
the physics of that transition, if it happens, is still uncertain
\citep{niemeyer1997b, niemeyer1999a}. Recent two-dimensional
calculations have suggested that burned material may quickly ascend to
the surface of the still gravitationally bound star, sweep around it
and, by collision and compression on the opposite side, trigger a
detonation in the unburned material \citep{plewa2004a}. This is called
by its proponents ``gravitationally confined detonation'', or GCD.

Here, we follow the evolution of one-sided ignitions for a range of
assumptions regarding the initial conditions in two-dimensional (2D) and
three-dimensional (3D) simulations. The results are sensitive to the
fuel consumption and energy release during the rising stage of the
plume. Therefore a correct description of the turbulent deflagration
flame is as crucial here as in other models. We describe our approach
in Sect.~\ref{sect:mod}.  Before that, we give a brief motivation of
the ignition scenarios explored here. Although \citet{roepke2006a}
showed that ignition conditions cannot be explored reliably in
2D simulations, we can and do use surveys in cylindrical
symmetry to get a feeling for the parameter range to be explored, as
well as the dependence of the results on numerical resolution
(Sect.~\ref{sect:2d}). Full-star 3D simulations are
presented in Sect.~\ref{sect:3d}, and the consequences regarding the
possibility of triggering a detonation are discussed in
Sect.~\ref{sect:sdet}. Conclusions are drawn in
Sect.~\ref{sect:concl}.

\section{Off-center ignition}

Several studies now suggest ignition with an offset from the center of
the WD of order $100\ldots 200 \, \mathrm{km}$
\citep{garcia1995a,woosley2004a,wunsch2004a,kuhlen2006a,iapichino2006a}.
It should be acknowledged that none of these studies has yet followed
the actual transition from a high temperature fluctuation to a flame
in a self-consistent way, including the possibility the perturbation
is disrupted by turbulence, and all fall far short of the actual
Reynolds number in the star.  Also, while arguments based upon a
probability density function can offer some guidance as to whether a
particular temperature is likely to be realized, they cannot, by
themselves, say whether the high temperature happens in a contiguous
region, or in many disparate points, or even over some interval of
time.

Still, the calculations of \citet{kuhlen2006a} do suggest that
ignition is unlikely to occur as a single spherical bubble either at
the center or off-center. Rather the distribution of high temperature
may look more like a ``teardrop'', spreading as it goes out to a large
opening angle.  We thus explore here a variety of initial conditions
ranging from nearly spherical bubbles far off center, to ignition in
multiple points forming complex configurations.  The models are named
according to their dimensionality and the ignition characteristics
where ``B'' stands for a single bubble (spherical as much as Cartesian
coordinates allow) with an attached number indicating its radius in
kilometers, ``P'' is a highly perturbed bubble,``T'' stands for a
teardrop distribution of multiple bubbles followed by a ``1'' to
indicate a strictly one-sided ignition and by a ``2'' when the
ignition region overshoots to the opposite side. Where appropriate,
``d'' followed by a number gives the distance of the ignition center
from the middle of the WD in the bubble case (and the maximum extent
of ignition measured from the WD's center in the teardrop case).  For
example, Model 3B50d200 is a 3D simulation with an ignition in form of
a spherical bubble of radius $50 \, \mathrm{km}$ centered 200 km from
the middle of the star. There may also be variations on these names
based upon resolution (a,b,c, etc).

\section{The astrophysical and numerical model}
\label{sect:mod}

The implementation of the deflagration \sn\ model follows the detailed
descriptions given by \citet{reinecke1999a,reinecke2002b,roepke2005c}
and \citet{schmidt2006c}. The hydrodynamics is modeled via the
piecewise parabolic method \citep{colella1984a} in the
\textsc{Prometheus} implementation \citep{fryxell1989a}, in combination
with a WD matter equation of state incorporating an electron gas
relativistic and degenerate to variable degrees, a Boltzmann gas of
nuclei, photons and electron-positron pairs.

Since the results are sensitive to small perturbations of the initial
flame configuration, the directional splitting in the hydrodynamics
solver, as well as discretization errors on the Cartesian grid may provide
seeds for developing instabilities. This is not necessarily
unphysical. In a realistic \sn\ explosion, the background is not
expected to be smooth, nor will the initial flame shape be perfectly
regular. But, as discussed below, one sometimes has little control
over these effects.

Another crucial aspect of the modeling is the prescription for flame
propagation. We strive to implement here a consistent model for
burning in the flamelet regime following standard theories of
turbulent combustion---such as they are. The fundamental assumption is
the establishment of a turbulent cascade from large-scale eddies
produced by shear instabilities at the interfaces of burning bubbles
down to the Kolmogorov scale, where viscous effects dissipate the
turbulent kinetic energy. The flame interacts with turbulence down to
the Gibson scale, where the laminar burning speed $s_\mathrm{l}$
becomes comparable to the turbulent velocity fluctuations $v'$. By
definition in the flamelet regime, this Gibson scale is large compared
to the width of the flame---a condition that holds for most parts of
the supernova explosion until quite low densities are reached
\citep[for an approach to modeling stages beyond this regime
see][]{roepke2005a}. Thus the internal flame structure remains
unaffected by turbulent eddies.  \citet{damkoehler1940a} first pointed
out that the turbulent flame front in these circumstances should
propagate with an effective velocity that is proportional to the
turbulent velocity fluctuations and independent of the burning
microphysics.

Following this concept, the flame is modeled as a sharp interface
separating the fuel from the ashes. Its propagation is followed in a
level set approach \citep{osher1988a}, where the flame front is
associated 
with the zero level set of a scalar, $G$, defined to be a signed
distance function away from the flame. This front is advanced as
described by \citet{reinecke1999a}.

The speed of flame propagation is a function of the turbulent velocity
fluctuations on the scale of the computational grid cells. Its value
is determined from a subgrid-scale turbulence model. In the
2D simulations, this model is implemented according to
\citet{niemeyer1995b}, and the turbulent flame speed, $s_\mathrm{t}$, is
set equal to the turbulent velocity fluctuations, $v'=\sqrt{2
  k_\mathrm{sgs}}$, where $k_\mathrm{sgs}$ denotes the subgrid-scale
turbulent specific kinetic energy. The 3D simulations
employ an improved subgrid-scale model
\citep{schmidt2006b,schmidt2006c}. That model, based upon localized closures
for the terms of the balance equation of turbulent subgrid-scale
energy, does not need to assume a specific scaling behavior of the
turbulent cascade, nor isotropy of the turbulence. The flame
propagation speed is implemented as
$$
s_\mathrm{t} = s_\mathrm{l} \sqrt{1 + C_\mathrm{t} \left(
    \frac{q_\mathrm{sgs}}{s_\mathrm{l}}\right)^2}
$$
with $C_\mathrm{t} = 4/3$ and $q_\mathrm{sgs}$ denoting the
subgrid-scale turbulence velocity \citep{pocheau1994a,schmidt2006c}.

Nuclear reactions are implemented in a simple way described by
\citet{reinecke2002b}. Only four species, $^{12}$C, $^{16}$O,
$^{24}$Mg, $^{56}$Ni, and $\alpha$-particles, are taken into
account. Material traversed by the flame front is converted to a
composition represented by a temperature-dependent mixture of
$^{56}$Ni and $\alpha$-particles in nuclear statistical equilibrium,
or, at fuel densities below $5.25 \times 10^7 \, \mathrm{g} \,
\mathrm{cm}^{-3}$, to intermediate mass elements represented by
$^{24}$Mg. At fuel densities below $10^7 \, \mathrm{g} \,
\mathrm{cm}^{-3}$ nuclear reactions are assumed to cease.

In all simulations the WD was set up cold and isothermal with a
temperature of $T=5\times 10^5 \, \mathrm{K}$ and a central density of
$\rho_\mathrm{c} = 2.9\times 10^9 \, \mathrm{g}\, \mathrm{cm}^{-3}$
composed of equal parts of carbon and oxygen.

The discretization on the computational grid follows the strategy of
two nested moving grids suggested by \citet{roepke2006a}, where a
fine-resolved uniform inner grid contains the flame while an outer
grid with exponentially growing grid cells accommodates the WD
star and follows its expansion. Due to flame propagation inside the
WD, it is possible to subsequently collect adjacent grid cells of the
outer grid into the uniform part, as soon as the cell sizes match
thereby optimizing the resolution for the given number of
computational grid cells.

\section{Exploring the possibilities: 2D simulations}
\label{sect:2d}

\begin{figure*}
\includegraphics[width = \textwidth]{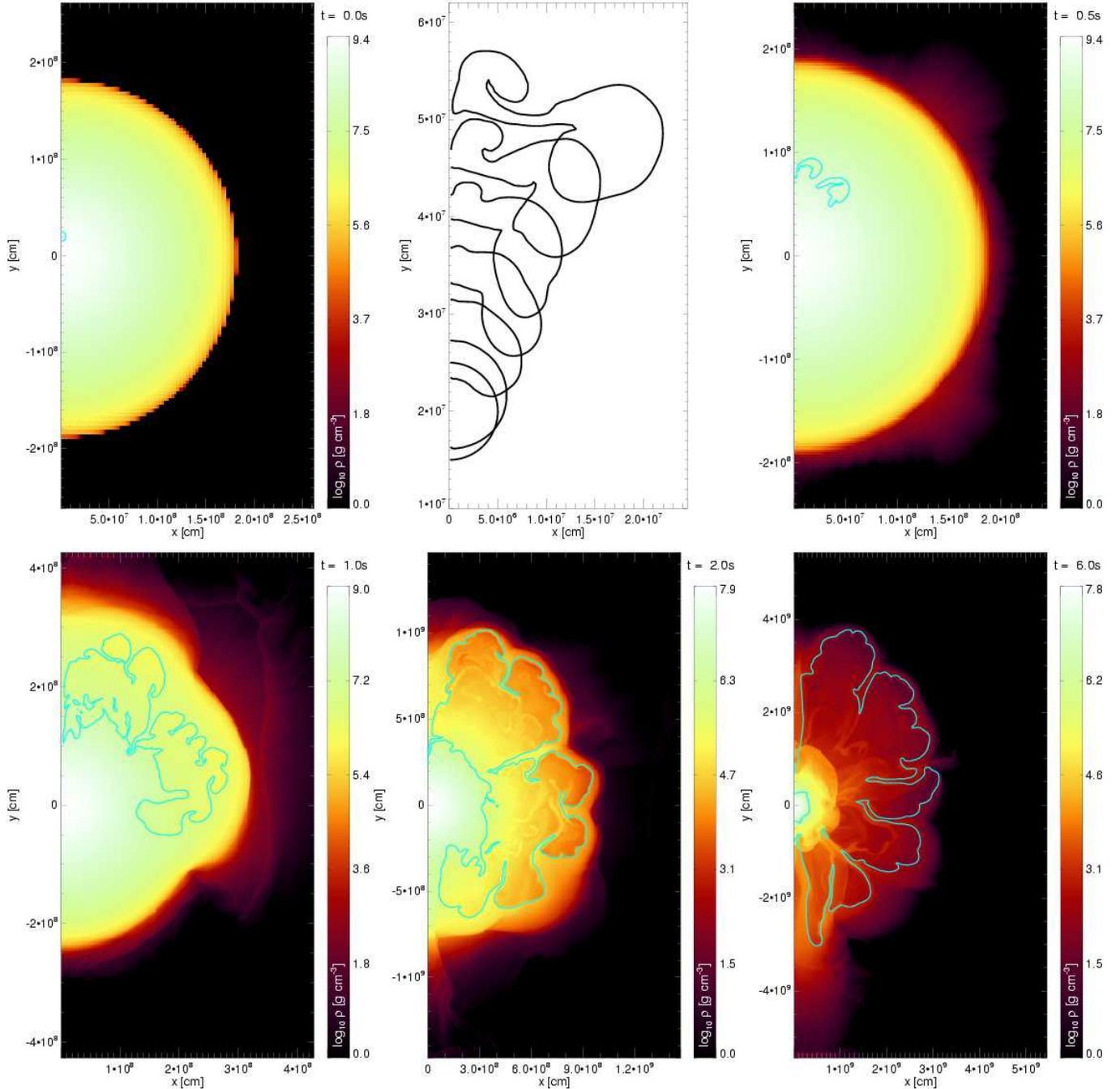}
\caption{Evolution of a 2D explosion simulation (Model 2B50d200c) ignited
  in a single bubble $200 \, \mathrm{km}$ off-center. The top-middle
  panel illustrates the flame front evolution in the interval $t =
  [0, 0.3]\,\mathrm{s}$. Each contour corresponds to a time step of
  $0.05 \, \mathrm{s}$. In all other snapshots, the cyan isosurface
  corresponds to the zero level set of $G$ which is
  associated to the flame front in early stages of the evolution and
  indicates the interface between fuel and ashes once burning has
  ceased. \label{fig:evo2d}}
\end{figure*}

The parameter space and the dependence of the results on numerical
resolution were first explored assuming cylindrical symmetry. The
results obtained in 2D should not be used to draw quantitative
conclusions, but are a numerically inexpensive way to explore a broad
range of possibilities.

In the cylindrical $(r,z)$-setups, the entire WD was accommodated on the
grid, from its center to its radius assuming rotational symmetry
about the $z$-axis (see top left panel of Fig.~\ref{fig:evo2d}).

\subsection{Single-bubble ignition}

As a first numerical experiment, the flame was initiated in the
simplest conceivable configuration: a single spherical bubble ignited
somewhere on the $z$-axis. Even this simple configuration has three
parameters that potentially impact the evolution of the explosion. One
is the resolution of the flame and the WD star. Two additional
parameters---the displacement of the igniting bubble from the center
of the WD and its radius---are more physical in nature. As an
illustrative model, consider one where ignition took place $200 \,
\mathrm{km}$ off-center in a spherical bubble of radius $50 \,
\mathrm{km}$ (cf.\ Fig.~\ref{fig:evo2d}). The resolution for this
example study (Model 2B50d200c, Table~\ref{tab:res}) was 256 grid
cells in $r$-direction and 512 grid cells in $z$-direction, and the
initial setup is shown in the top left panel of
Fig.~\ref{fig:evo2d}. In the top middle panel, a close-up on the flame
illustrates the evolution from the initial spherical shape to a more
irregular toroidal structure. This is a consequence of the
buoyancy-induced flotation of the bubble acting in combination with
self-propagation of the flame due to burning. The flame floats towards
the surface of the star, thereby being subject to considerable lateral
spread (cf.\ snapshots at $t = 0.5 \, \mathrm{s}$ and $t = 1.0 \,
\mathrm{s}$ in Fig.~\ref{fig:evo2d}). At $t \sim 2 \, \mathrm{s}$,
burning has ceased since the fuel density ahead of the flame has
dropped below $10^7 \, \mathrm{g} \, \mathrm{cm}^{-3}$. The burned
material sweeps around the star since it is still gravitationally
bound. Although it extends to rather large radii, only about $0.49 \,
M_\odot$ is located outside a radius of $2.5 \times 10^8 \,
\mathrm{cm}$ at $t = 2 \, \mathrm{s}$. Only $\sim 0.1 \, M_\odot$ of
the WD is burned, inadequate to unbind the star. In the outer layers
of the WD opposite to the flame ignition, burned material collides
subsonically with a compressional front moving ahead of the actual
ash.  In the simulation, the evolution continues by an expansion of
the outer layers while the central parts of the WD contract such that
the inner parts of the flame eventually reach fuel densities above the
threshold of $10^7 \, \mathrm{g} \, \mathrm{cm}^{-3}$. In our flame
description burning resumes at this point, and the newly processed
material is expelled into the ash region (cf.\ Fig.~\ref{fig:evo2d},
snapshot at $t = 6.0 \, \mathrm{s}$). This, however, is sensitive to
the way burning is implemented in the code (either completely ``on''
for densities above 10$^7$ g cm$^{-3}$, or  ``off'' for lower
densities) and may not be a realistic occurrence.

\begin{figure}
\includegraphics[width = 0.7 \linewidth]{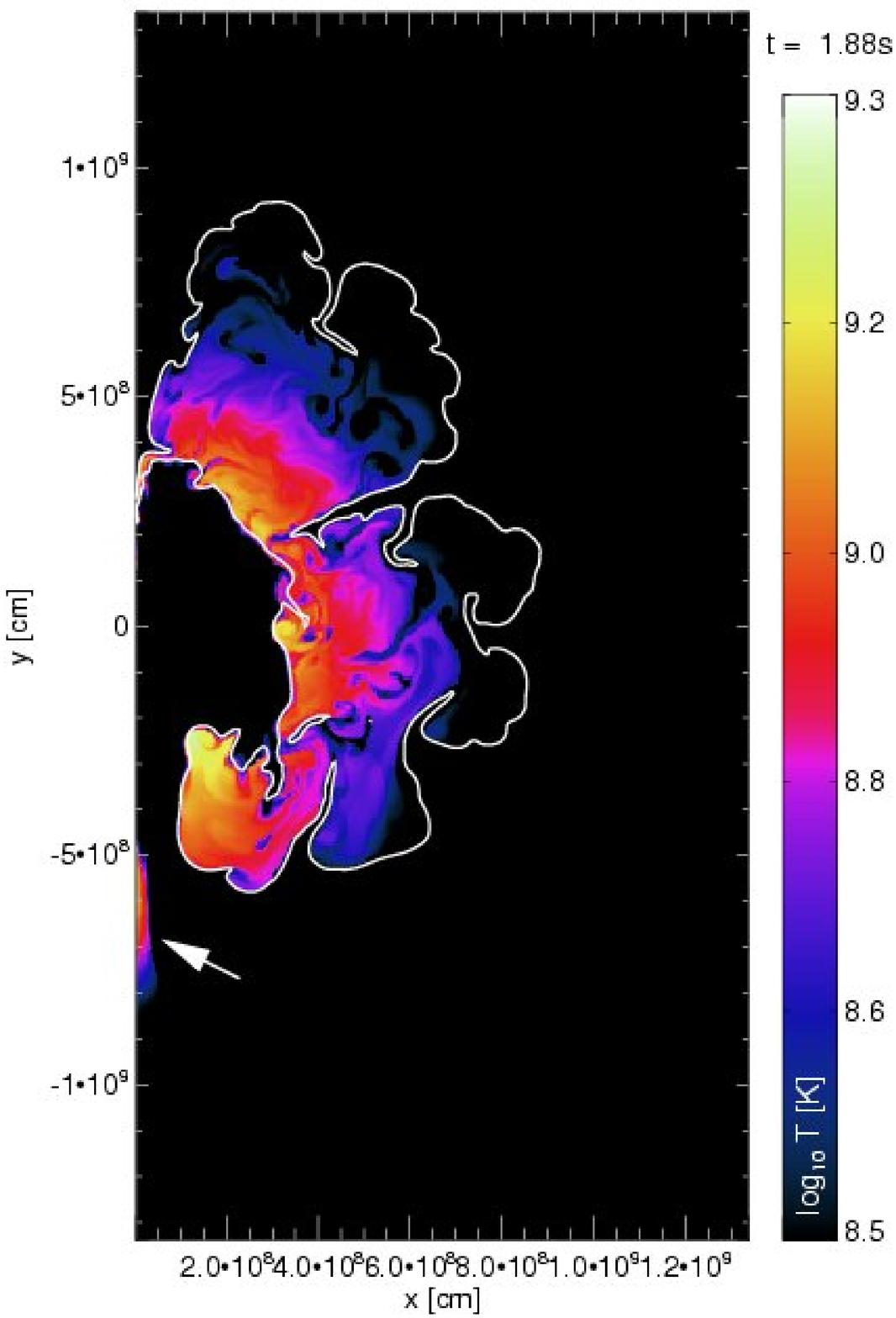}
\caption{Temperature distribution Model 2B50d200c. The white contour
  indicates the zero-level set of 
  $G$ indicating the interface between fuel and ashes. Apart from the
  ash regions enclosed by this contour, significantly
  increased temperatures are found in the region
  compressed by the collision (marked by the white arrow).\label{fig:coll_tem}}
\end{figure}

As pointed out by \citet{plewa2004a}, the collision of the
gravitationally bound material sweeping around the surface of the WD
marks an interesting point in the evolution.  To initiate a
spontaneous detonation, high density and temperature are both
necessary in the unburned material at the focus of the collision. The
cylindrical-symmetric setup forces the collision of burned material to
take place at the negative part of the $z$-axis.
Figure~\ref{fig:coll_tem} shows the temperature of the material in the
model described above. The snapshot was taken at the instant of peak
temperature in the compressed fuel in the collision region. The peak
temperature there was $2.22 \times 10^9 \, \mathrm{K}$ at a density of
$1.41 \times 10^6 \, \mathrm{g} \, \mathrm{cm}^{-3}$.  As will be
discussed, the conditions reached in the collision depend on the
parameters of the setup as discussed.

\subsubsection{A note on numerical convergence}

\begin{table*}
\begin{center}
\caption{2D resolution study for ignition $200 \, \mathrm{km}$ off-center
  in a bubble of $50 \, \mathrm{km}$ radius. Dashes mark cases where
  the density at which to measure the temperature is not reached in
  the collision.
\label{tab:res}}
\setlength{\extrarowheight}{2pt}
\begin{tabular}{p{6em}lllllll}
\tableline\tableline
Model&
\multicolumn{1}{p{4.6em}}{bubble\newline radius [km]} &
\multicolumn{1}{p{5em}}{resolution} &
\multicolumn{1}{p{4.85em}}{$T_\mathrm{max}$ at coll.\ [$10^9 \, \mathrm{K}$]} &
\multicolumn{1}{p{4.67em}}{$E_\mathrm{nuc}$ at coll.\ [$10^{50} \, \mathrm{erg}$]}&
\multicolumn{1}{p{9.9em}}{$T_\mathrm{max} (\rho > 3 \times
  10^6 \, \mathrm{g} \, \mathrm{cm^{-3}})$ at coll.\ [$10^9 \, \mathrm{K}$]}&
\multicolumn{1}{p{9.9em}}{$T_\mathrm{max} (\rho > 1 \times
  10^7 \, \mathrm{g} \, \mathrm{cm^{-3}})$ at coll.\ [$10^9 \,
  \mathrm{K}$]} &
\multicolumn{1}{p{4.5em}}{surface\newline detonation?}\\
\tableline
2B50d200a & 50 & $128 \times 256$   & 2.61 & 1.14 & 1.54  & ---  & no \\
2B50d200b & 50 & $192 \times 384$   & 2.92 & 0.97 & 2.60  & ---  & yes \\
2B50d200c & 50 & $256 \times 512$   & 2.22 & 1.46 & 1.28  & --- & no \\
2B50d200d & 50 & $384 \times 768$   & 2.53 & 1.44 & 0.959 & --- & no \\
2B50d200e & 50 & $512 \times 1024$  & 2.29 & 1.39 & 0.954 & --- & no \\
\tableline
2B25d200a & 25 & $128 \times 256$   & 2.40 & 1.33 & 2.08  & ---  & no \\
2B25d200a & 25 & $192 \times 384$   & 1.97 & 1.47 & 0.224 & ---  & no \\
2B25d200a & 25 & $256 \times 512$   & 2.60 & 1.09 & 2.32  & --- & yes \\
2B25d200a & 25 & $384 \times 768$   & 3.03 & 0.72 & 3.03  & 2.95  & yes \\
2B25d200a & 25 & $512 \times 1024$  & 3.83 & 0.82 & 3.83  & 3.80   & yes \\
\tableline
\end{tabular}
\end{center}
\end{table*}

Two sets of simulations with spherical initial flames were carried out
varying the number of cells in the computational grid. One had an
initial bubble radius of $50 \, \mathrm{km}$, and the other, a bubble
radius of $25 \, \mathrm{km}$. For each simulation, the maximum
temperature in the unburned material in the collision region was
determined, along with the nuclear energy release prior to reaching
this temperature. The results are given in Table~\ref{tab:res}. There
is significant scatter in the critical collision temperature with
values deviating by up to 43\%. A similar variation is seen in the
energy of the 
burning. Although a clear trend is not apparent in
Table~\ref{tab:res}, it seems that better numerical resolution leads
to less energy release in the burning, less expansion, and hence to a
stronger collision in the unburned material. The scatter is larger
when starting with bubbles of smaller radius.

Previous studies suggest convergent results should be obtained with a
resolution of about 256 cells per dimension in one octant \citep{reinecke2002b,
  roepke2005c}. Here the situation is different. While previous
resolution studies were carried out on the basis of an initial flame
setup with well specified perturbations imposed on it, all seeds for
growing nonlinear instabilities in the spherical bubble setup applied here are
introduced by numerical artifacts, such as discretization errors and
noise. This setup is not a well-posed numerical problem because the
evolution of nonlinear features in the flame structure is expected to
be strongly resolution dependent and thus
numerical convergence is problematic.  Given that the flame evolution
is dominated by strongly nonlinear effects, the variation in Table 1
is no big surprise and is illustrative of the uncertainty in our
results. We emphasize though that this is not due to the numerical
methods applied here, but due to the variable (and artificial) initial
setup.

Viewed this way, the results of the resolution study can be understood
in a straightforward manner. 
It should be noted that the resolution affects the answer in two
ways. On the one hand, it affects the temperature in the collision because
higher resolution smears out the hot spot less. But, on the other hand, it also
affects the propagation of the flame as it moves
outwards in the star, as illustrated by the variable explosion
energy. Higher resolution decreases the
discretization errors and therefore reduces the seeds for the growth
of nonlinear perturbations. Therefore the flame develops less surface,
less material is consumed and the lower energy release leads to a
weaker expansion of the star. The material sweeping around at the
surface is stronger gravitationally bound and the clash is more
vigorous.  Smaller initial flame bubbles are more sensitive to the
numerical resolution. This is understandable since here bubble
flotation is slower and nonlinear features invoked by discretization
errors have more time to develop on the way to the surface of the WD.

An imaginary ideal situation with no discretization errors would
suppress the nonlinear growth due to instabilities. But since the WD
star is expected to be perturbed by pre-ignition convection and the
flame is likely to ignite in multiple spots or an irregular shape,
this seems far from reality.

Besides the energy release due to burning, the impact of the
colliding material will also be sensitive to the morphology of the
ash that is driving the unburned material like a piston. A well-defined,
large leading edge of the colliding ash regions should result
in a better focus than a multitude of leading features. Since the
flame morphology is determined by nonlinear effects and instabilities,
discretization errors in different resolutions amplify
the scatter in the results.

\subsubsection{Bubble displacement}

\begin{figure}
\includegraphics[width = \linewidth]{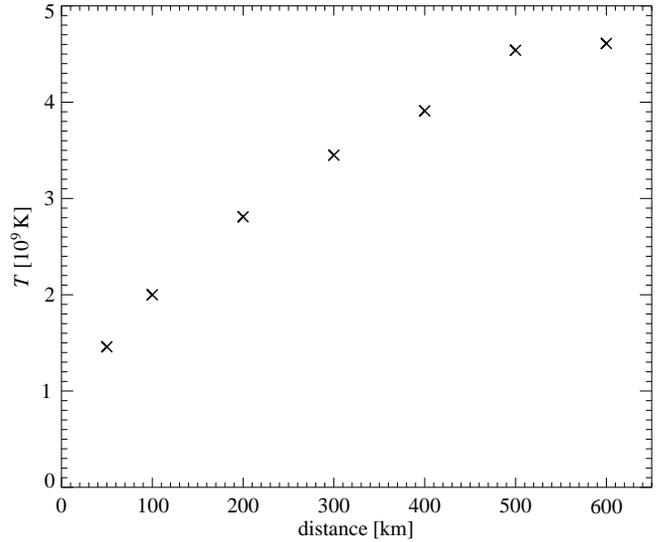}
\caption{ Maximum fuel temperature reached in the region of
  collision of the material sweeping around the WD as a function of
  the displacement of the initial flame bubble from the center of the
  WD (according to Table~\ref{tab:radvar}).\label{fig:radvar}}
\end{figure}

The first physical parameter---the displacement of the initial flame
bubble from the center---turned out to have substantial influence on
the densities and temperatures reached in the collision. Values from
simulations starting with an initial flame bubble of radius $25 \,
\mathrm{km}$ and varying distances from the center of the WD are given
in Table~\ref{tab:radvar}. In order to minimize the scatter due to
differences in resolution, the initial grid spacing of the $256\times
512$ cells setup was held fixed for all simulations. Note, however,
that since two nested grids are used to follow the expansion of the WD
and the flame propagation, the resolution evolves according to the
energy released in the burning, which is different in the various
simulations. Nevertheless, a trend of increasing collision
temperatures with larger initial flame displacements from the center
of the WD is clearly visible (cf.\ Fig.~\ref{fig:radvar}). At large
displacements the temperature increases less indicating a saturation
of the effect. A possible explanation is that although here the
expansion of the star prior to breakout is decreased, the amount of
ashes expelled from the surface also decreases due to less burning
taking place. Therefore the momentum of the colliding ash regions is
smaller and the reduced impact leads to lower compression
temperatures.  Flames born closer to the center burn more material and
cause more expansion (cf.\ Table~\ref{tab:radvar}). Displacing the
initial flame bubble from $50 \, \mathrm{km}$ to $600 \, \mathrm{km}$
off-center decreases the nuclear energy release for 94\% and increases
the maximum temperature reached in the collision by 250\%. Such a
large displacement as 600 km is not realistic, but the consequences
may be the same as for a bubble ignited closer in, but with a less
efficient prescription for burning on the way out.

\begin{table*}
\begin{center}
\caption{Collision parameters in dependence of initial bubble
  displacement (2D simulations; initial bubble radius: $25 \,
  \mathrm{km}$). Dashes mark cases where
  the density at which to measure the temperature is not reached in
  the collision.
\label{tab:radvar}}
\setlength{\extrarowheight}{2pt}
\begin{tabular}{p{6em}llllll}
\tableline\tableline
Model &
\multicolumn{1}{p{5.5em}}{distance from center [km]} &
\multicolumn{1}{p{4.85em}}{$T_\mathrm{max}$ at coll.\
  [$10^9 \, \mathrm{K}$]} &
\multicolumn{1}{p{4.67em}}{$E_\mathrm{nuc}$ at coll.\ [$10^{50} \, \mathrm{erg}$]}&
\multicolumn{1}{p{9.9em}}{$T_\mathrm{max} (\rho > 3 \times
  10^6 \, \mathrm{g} \, \mathrm{cm^{-3}})$ at coll.\ [$10^9 \, \mathrm{K}$]}&
\multicolumn{1}{p{9.9em}}{$T_\mathrm{max} (\rho > 1 \times
  10^7 \, \mathrm{g} \, \mathrm{cm^{-3}})$ at coll.\ [$10^9 \,
  \mathrm{K}$]} &
\multicolumn{1}{p{4.5em}}{surface\newline detonation?}\\
\tableline
2B50d50  & 50  & 1.46 & 2.31 & 0.809 & --- & no\\
2B50d100 & 100 & 2.00 & 1.93 & --- & --- & no\\
2B25d200 & 200 & 2.81 & 1.07 & 2.13  & --- & no\\
2B25d300 & 300 & 3.45 & 0.76 & 3.45  & 0.438  & yes\\
2B25d400 & 400 & 3.91 & 0.46 & 3.91  & 3.91   & yes\\
2B25d500 & 500 & 4.54 & 0.28 & 4.54  & 3.91   & yes\\
2B25d600 & 600 & 4.61 & 0.12 & 4.61  & 4.61   & yes\\
\tableline
\end{tabular}
\end{center}
\end{table*}

\subsubsection{Bubble morphology}

\begin{figure}
\centerline{\includegraphics[width = 0.9 \linewidth]{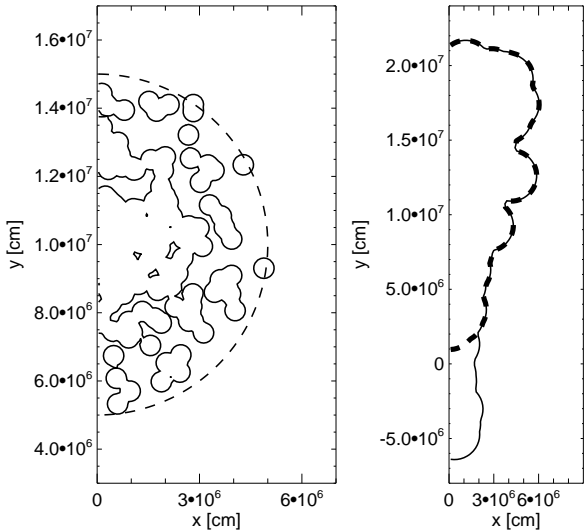}}
\caption{Different initial flame bubble morphologies. Left: perfect sphere
  (dashed) vs.\ irregular shape (solid); right: teardrop-shaped
  initial flames on one side of the WD's center (dashed) and
  overshooting to the opposite side of it (solid).\label{fig:morph}}
\end{figure}

\begin{table*}
\begin{center}
\caption{Collision parameters for 2D simulations with different
  ignition morphologies. Dashes mark cases where
  the density at which to measure the temperature is not reached in
  the collision.
\label{tab:morph}}
\setlength{\extrarowheight}{2pt}
\begin{tabular}{p{6em}llllll}
\tableline\tableline
Model &
\multicolumn{1}{p{5em}}{resolution} &
\multicolumn{1}{p{4.85em}}{$T_\mathrm{max}$ at coll.\
  [$10^9 \, \mathrm{K}$]} &
\multicolumn{1}{p{4.67em}}{$E_\mathrm{nuc}$ at coll.\ [$10^{50} \, \mathrm{erg}$]}&
\multicolumn{1}{p{9.9em}}{$T_\mathrm{max} (\rho > 3 \times
  10^6 \, \mathrm{g} \, \mathrm{cm^{-3}})$ at coll.\ [$10^9 \, \mathrm{K}$]}&
\multicolumn{1}{p{9.9em}}{$T_\mathrm{max} (\rho > 1 \times
  10^7 \, \mathrm{g} \, \mathrm{cm^{-3}})$ at coll.\ [$10^9 \,
  \mathrm{K}$]} &
\multicolumn{1}{p{4.5em}}{surface\newline detonation?}\\
\tableline
2P50d100 & $384\times 768$ & 2.31 & 1.14 & 2.31 & 0.507 & yes\\
2B50d100 & $384\times 768$ & 4.25 & 0.861 & 4.25 & 4.25 & yes\\
2T1d200  & $256\times 512$ & 2.21 & 1.25 & 0.991  & --- & no\\
2T2d200  & $256\times 512$ & 0.958 & 2.30 & 0.0797  & ---  & no\\
\tableline
\end{tabular}
\end{center}
\end{table*}

The initial morphology of the flame also affects the strength of the
collision. Any divergence from a perfectly spherical shape has a similar
effect to varying the resolution. In both cases, seeds for the
developing instabilities are imposed---with changing resolution due
to discretization errors, and for more complex initial bubble shapes,
explicitly in a controlled way.

To demonstrate this, two simulations were carried out on a well
resolved ($384\times768$ cells) computational grid. One model
(2B50d100, see Table~\ref{tab:morph}) was ignited in a---within
discretization error---perfectly spherical bubble of radius $50 \,
\mathrm{km}$ at a distance of $100 \, \mathrm{km}$ from the center. In
the second simulation (2P50d100), the initial flame was composed of
160 partially overlapping small bubbles of radius $3 \, \mathrm{km}$
placed within a sphere of $50 \, \mathrm{km}$ radius $100 \,
\mathrm{km}$ off-center of the WD (cf.\ Fig.~\ref{fig:morph}, left).

While the spherical initial flame led to a maximum
temperature of $4.25 \times 10^9 \, \mathrm{K}$ in the collision of
the surface material and released $0.861 \times 10^{50} \,
\mathrm{erg}$ prior to the collision, the irregular-shaped initial
flame caused much more burning. It released $1.14 \times 10^{50} \,
\mathrm{erg}$ before the clash, and the maximum temperature reached
in the collision region was only $3.31 \times 10^9 \, \mathrm{K}$
(cf.\ Table~\ref{tab:morph}).

\subsubsection{Irregular asymmetric ignition}

\begin{figure*}
\centerline{\includegraphics[width = \linewidth]{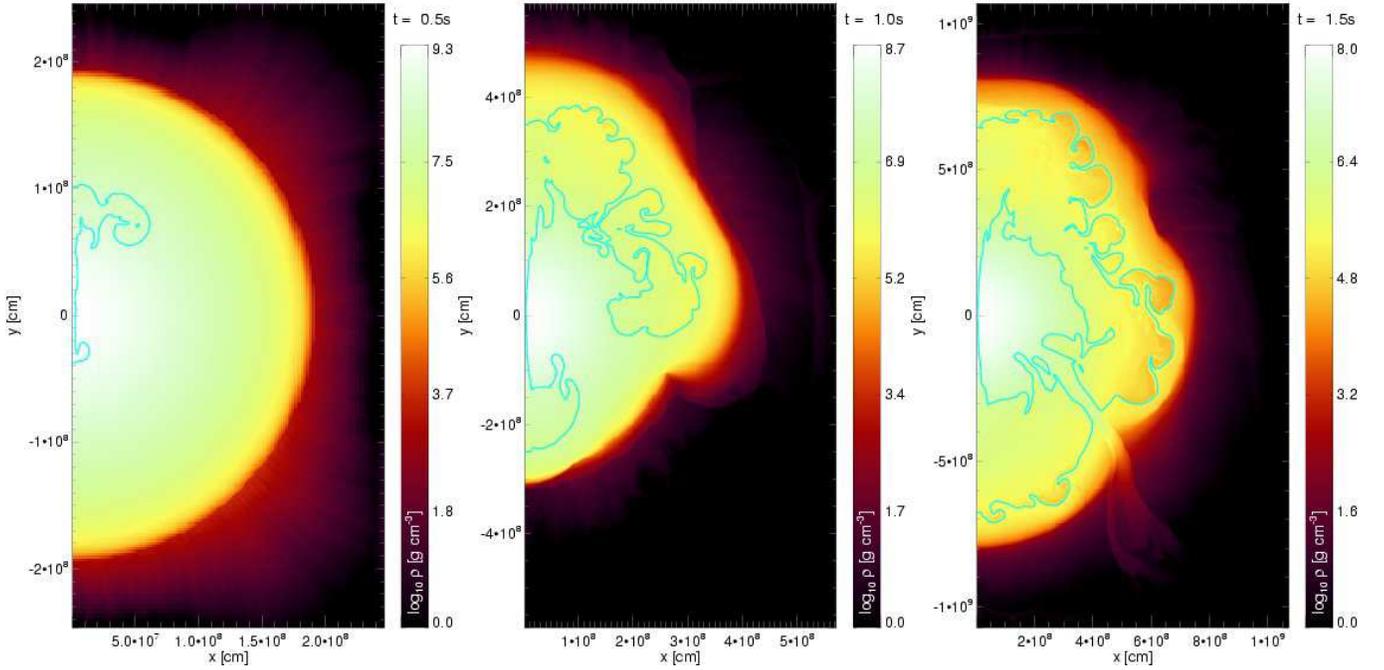}}
\caption{Evolution of an explosion simulation with the flame ignition
  extending to opposite sides of the WD (Model 2T2d200).\label{fig:td}}
\end{figure*}

Due to the dipolar convection flow structure found in the pre-ignition
phase by \citet{kuhlen2006a}, a lop-sided flame ignition---possibly
in many separate spots---is plausible. That ignition region may extend
out to $\sim$$200 \, \mathrm{km}$ and reach down to the center of the
WD. Some overshooting to the opposite side may also be possible if the
actual flow is multipolar. The ignition kernels will quickly merge due
to burning and the resulting flame structure may look like a teardrop
with an irregular surface.

We examined two such configurations, one ignited on only one side of
the WD (Model 2T1d100, cf.\ Table~\ref{tab:morph}), and another in
which the ignition extended through the center (Model 2T2d100).  These
ignition configurations are shown in the right hand plot of
Fig.~\ref{fig:morph}.

While the one-sided ignition evolution proceeded in a way similar to
the single-bubble ignition simulations (Model 2B50d200c of
Table~\ref{tab:res} can serve as a reference simulation; its evolution
is shown in Fig.~\ref{fig:evo2d}), the extension of the ignition
region to the opposite side of the WD's center had dramatic
consequences. The burning did not ascend to the surface of the star on
just one side, but evolved into two large irregular bubbles that moved
in both directions (cf.\ Fig.~\ref{fig:td}). Naturally, the collision
of material sweeping around the surface occurred off-axis. In this
set-up, two opposing effects altered the collision strength. On the
one hand, burning material on both sides of the star releases more
energy and therefore the expansion of the star proceeds more
rapidly. But, on the other hand, the collision occurs only slightly
more than half way around the hemisphere, and therefore takes place
earlier.

The reference Model 2B50d200c reached a collision temperature of $2.22
\times 10^9 \, \mathrm{K}$  at a
density of $1.41 \times 10^6 \, \mathrm{g} \, \mathrm{cm}^{-3}$ and
released $1.46 \times 10^{50} \,
\mathrm{erg}$ in the burning. The one-sided teardrop ignition led to
similar values---a maximum temperature of $2.21 \times 10^9 \,
\mathrm{K}$ at a density of $3.57 \, \times 10^6 \, \mathrm{g} \,
\mathrm{cm}^{-3}$  and a nuclear energy release of $1.25 \times 10^{50} \,
\mathrm{erg}$.

For the two-sided teardrop ignition, considerably more energy ($2.30
\times 10^{50} \, \mathrm{erg}$) was released in the burning prior to
the collision. This decreased the collision strength dramatically.
The temperature of the compressed fuel did not exceed $10^9 \,
\mathrm{K}$ (cf.\ Table~\ref{tab:morph}), indicating that the effect
of the earlier clash of the material coming from both poles cannot
compensate for the enhanced expansion.  However, it cannot be ruled
out that multiple plumes breaking out at smaller angles with the
center of the WD may collide more efficiently. But in this case the
initial conditions have to be chosen carefully since bubbles too close
to each other will merge before reaching the surface. Looking at the
evolution of our models ignited in a single bubble, this seems rather
hard to achieve. The snapshot at $t=1.0\,\mathrm{s}$ of Model
2B50d200c shown in fig.~\ref{fig:evo2d} indicates that by the time the
burnt structure breaks our of the surface of the star, it spans an
opening angle of about $180^{\circ}$ with the WD's center making the
opposite-sided ignition the best choice. 

\begin{figure}
\includegraphics[width=\linewidth]{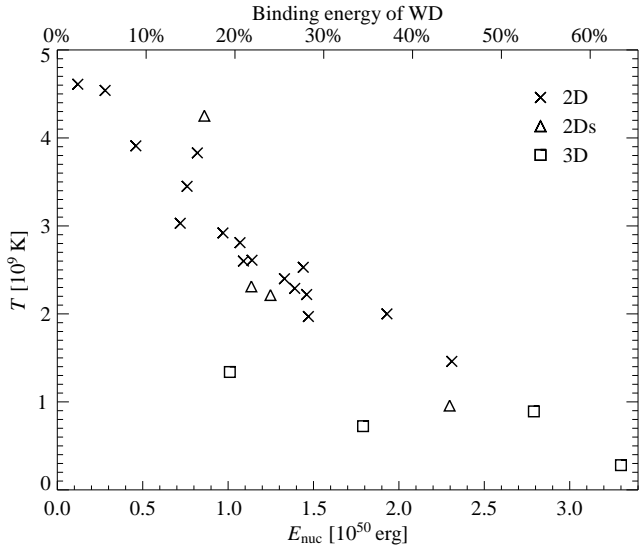}
\caption{\label{fig:et1} Maximum fuel temperature reached in the
  collision region as a function of
  the nuclear energy release prior to reaching that temperature 
  according to Tables~\ref{tab:res}, \ref{tab:radvar} (data points
  marked ``2D''), Table~\ref{tab:morph} (data points marked ``2Ds'')
  and Table~\ref{tab:3d} (data points marked ``3D'').}
\end{figure}

\subsection{Lessons learned from 2D simulations}

Our exploration of setup parameters in 2D simulations
reveals that a key quantity determining the collision strength is the
nuclear energy released on the flame bubble's way to the
surface. Combining the data of Tables~\ref{tab:res}, \ref{tab:radvar},
and \ref{tab:morph}, a clear correlation is visible between the
maximum temperature reached in the collision of surface material and
the amount of burning (cf.\ Fig.~\ref{fig:et1}).  Whether accomplished
through a change in resolution, a displacement of ignition point, or a
different morphology for the ignition region, less burning on the way
out correlates with a stronger, hotter collision on the far side.
This correlation arises naturally because the expansion of the star
leads to the collision being spread out over a larger volume. The lower
density also implies a greater heat capacity in the radiation
field. On the other hand, more burning also implies more ash
participating in the collision, which might make it stronger. But
apparently the expansion effect dominates.

\begin{figure*}
\centerline{\includegraphics[width = 0.98 \linewidth]{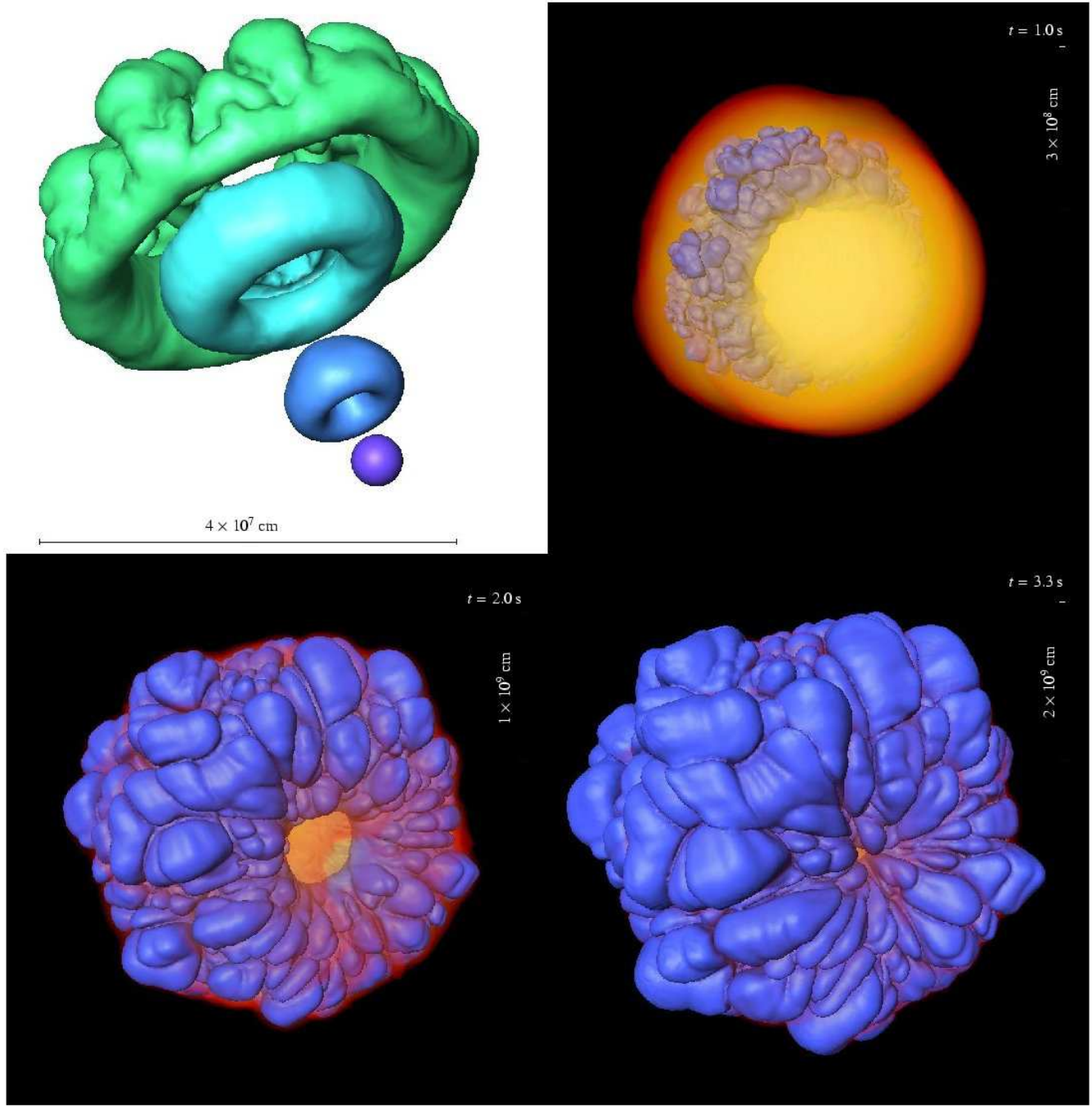}}
\caption{Evolution of a 3D explosion simulation with the flame ignition
 in a single bubble of radius $25 \, \mathrm{km}$ displaced $100 \,
 \mathrm{km}$ from the center of the WD (Model 3B25d100). Top left
 panel: initial 
 evolution of the flame front (blue to green isosurfaces correspond to
 $t = [0, 0.25, 0.35, 0.45] \, \mathrm{s}$). Other panels: later
 evolution with the 
 logarithm of the density volume rendered and $G=0$ as blue isosurface
 indicating the flame front or, later, the approximate boundary
 between burned and unburned material. \label{fig:evo3d}}
\end{figure*}

A potential concern is that insufficient resolution of the compressed
fuel region might make the temperature measurements
unreliable. However, the clear correlation shown in Fig.~\ref{fig:et1}
originates from simulations with different resolutions. The data
points lining up well is an indication that the temperature measurement was
credible even in the less resolved simulations

The amount of fuel burned is a consequence of many uncertain
aspects of the explosion physics, the specific algorithm
used to implement the flame propagation, and the resolution of the
simulation. For displacements that are not too extreme, stronger
collisions are favored by increased
distance of the ignition from the center.  Alterations of the
initial bubble shape diverging from the idealized spherical bubble
model also have a substantial impact on the flame propagation. More
complex initial flame shapes provide seeds for the developing
nonlinear flame features.

Two-sided ignition naturally burns more material, leading to
greater expansion. This is only partially compensated by the earlier
collision time.

\section{The full story: 3D simulations}
\label{sect:3d}

The entire WD was mapped onto a \emph{Cartesian} computational grid
and, again, different ignition setups were tested.  Compared with 2D,
several general factors alter the collision strength in 3D. In the
2D-simulations, each flame feature corresponds to a torus extending
around the star. Such complete burning does not occur in
three dimensions, and so one might expect stronger collisions in 3D due
to decreased expansion. On the other hand, the additional degree of
freedom also enhances the growth of the flame surface due to
instabilities. This effect has been seen in previous simulations where
3D-, centrally-ignited setups released significantly more energy than
their 2D-counterparts \citep{reinecke2002b, roepke2005c}. This causes
more expansion and weaker collisions.

Finally, when the collision geometry is no longer restricted by
cylindrical geometry, one expects less focusing, again weakening the
collision strength.

\subsection{Single-bubble ignition}

In two of the simulations, the flame was again ignited as a single
spherical bubble. The subsequent evolution is given in
Fig.~\ref{fig:evo3d}. Due to the interplay of burning and buoyancy,
the burning bubble alters its shape from a sphere to a torus during
the first few tenths of a second (cf., upper left panel of
Fig.~\ref{fig:evo3d}).  This evolution is very similar to the results
of \citet{zingale2005b} who simulated a burning bubble on small scales
fully resolving the flame structure.
The developing torus is more regular than the
respective flame structures found in the 2D
simulations. Nonetheless, it is subject to instabilities, and irregular
features grow, mostly on the outward side of the burning
region. The seeds for these features are probably discretization
errors. 

\begin{figure}
\centerline{\includegraphics[width = \linewidth]{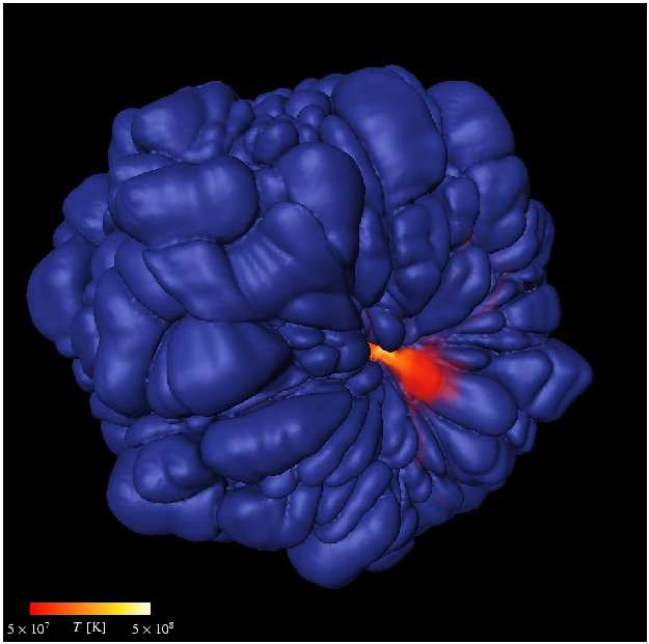}}
\caption{Snapshot of Model 3B25d100 at $t = 3.3 \, \mathrm{s}$, as in
  Fig.~\ref{fig:evo3d}, but here with the temperature volume rendered
  instead of the density.\label{fig:tem3d}}
\end{figure}

Once the ashes reach the outer parts of the star (n.b., not
necessarily the surface), they start to sweep around its core.
Interestingly, the leading edge of this sweeping material is defined,
even in 3D, by the former torus and therefore only slightly irregular
(cf.\ upper right and lower left panels of
Fig.~\ref{fig:evo3d}). Consequently, the clash of the burned material
still takes place in a well-defined spot on the opposite side of the
WD (cf.\ lower right panel of Fig.~\ref{fig:evo3d}). In comparison
with 2D simulations, this effect partially compensates
for the lack of symmetry restrictions and makes the focus of the
collision sharper than expected. In the collision, the temperature
increases, as shown by the volume rendering of the temperature field
in Fig.~\ref{fig:tem3d}.

In the two simulations presented here, the initial flame bubbles were
displaced $100 \, \mathrm{km}$ (Model 3B25d100) and $200 \,
\mathrm{km}$ (Model 3B25d200) from the center of the WD,
respectively. Both simulations were carried out on a $[512]^3$ cells
computational grid.  To gain the maximum possible resolution, the
fine-spaced inner uniform part of the computational grid was extended
only slightly beyond the ignition radius, and therefore the initial
resolution of the flame was coarser in the model ignited further
off-center. Contrary to the 2D study where the initial
resolution was kept constant while varying the ignition position, we
had to sacrifice comparability between the simulations to better
resolution in the model ignited closer to the center.  Because of the
computational expense of 3D simulations, 
two parameters were changed at the same time, i.e., the distance of the
ignition from the center and the perturbation imposed on the igniting
bubble due to discretization errors.

\begin{figure*}
\centerline{\includegraphics[width = \linewidth]{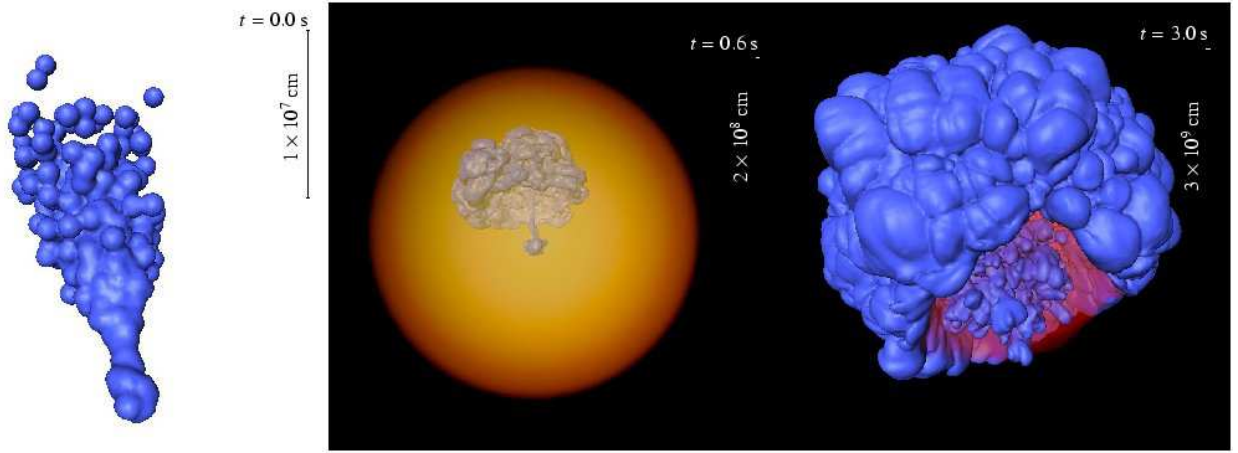}}
\caption{Evolution of an explosion simulation with the flame ignition
  in a two-sided teardrop shape (Model 3T2d200). The first snapshot
  shows the initial 
  flame (blue isosurface) with parts extending to both sides of the
  center of the WD. This leads to a two-sided evolution of the flame
  inside the WD, whose extend is indicated by the volume rendering of
  the logarithm 
  of the density (center panel). The energy released in the
  thermonuclear burning is sufficient to overcome the gravitational
  binding of the star, giving rise to a weak explosion. In the right
  snapshot, the blue isosurface now indicates the approximate
  interface between burned and unburned material. This configuration
  is approaching homologous expansion and no collision between ash
  regions will take place at the surface of the
  star. \label{fig:td3d}}
\end{figure*}

Unlike the 2D results, the flame ignited at $200 \,
\mathrm{km}$ off-center released more energy in 3D than the one
ignited at $100 \, \mathrm{km}$ off-center. The reason, most likely,
is the different discretization errors. The large features that
develop at the end of the torus-dominated phase of the evolution
were more pronounced in Model 3B25d200. Therefore more material was
consumed and the energy release of $5.65 \times 10^{50}\,\mathrm{erg}$
was even sufficient to unbind the star (which has a binding energy of
$-5.20 \times 10^{50}\,\mathrm{erg}$). In the simulation ignited $100 \,
\mathrm{km}$ off-center, the WD remained bound and the ashes breaking out
of the surface swept around the core and collided on the opposite side
as illustrated in Fig.~\ref{fig:evo3d}.

\begin{table}
\begin{center}
\caption{Collision parameters in the 3D simulations.
\label{tab:3d}}
\setlength{\extrarowheight}{2pt}
\begin{tabular}{p{5.5em}llll}
\tableline\tableline
Model &
\multicolumn{1}{p{4.85em}}{$T_\mathrm{max}$ at coll.\
  [$10^9 \, \mathrm{K}$]} &
\multicolumn{1}{p{4.67em}}{$E_\mathrm{nuc}$ at coll.\ [$10^{50} \, \mathrm{erg}$]}&
\multicolumn{1}{p{4.55em}}{$\rho$ at coll.\ [$10^6 \, \mathrm{g} \, \mathrm{cm^{-3}}$]}\\
\tableline
3B25d100 & 0.892 & 2.79 & $< 1.3 \times 10^5$ \\
3P25d100 & 1.34  & 1.01 & $< 1.5 \times 10^5$ \\
3P50d100 & 0.724 & 1.79 & $< 2.5 \times 10^5$ \\
3B25d200 & \multicolumn{3}{c}{no collision: WD unbound}\\
3T1d200   & 0.281 & 3.30 & $< 3.2 \times 10^3$ \\
3T2d200& \multicolumn{3}{c}{no collision: WD unbound}\\
\tableline
\end{tabular}
\end{center}
\end{table}

To gain more control over the initial perturbations that later affect
the growth and buoyancy of the flame, two additional simulations were
carried out where the ignition was, on the average, spherical, but
actually composed of many smaller spheres of hot ash, so that the
seeds for instabilities were present from the beginning. One, Model
3P25d100, had parameters similar to Model 3B25d100, with the center of
the ignition region positioned $100 \, \mathrm{km}$ off-center, and
the small flame kernels located inside a radius of $25 \,
\mathrm{km}$. For the second, Model 3P50d100, the displacement of the
center of the flame was unchanged, but the spherical aggregate of
small bubbles filled a larger volume with radius of $50 \,
\mathrm{km}$. In contrast to the 2D simulation, the
perturbations applied here decreased the amount of burning
(Table~\ref{tab:3d}). This can be understood from the the temporal
evolution of the energy release and the bubble morphology. In the
beginning, the energy release in the highly perturbed case exceeds
that for a smooth sphere, as expected from the faster development of
its flame surface. However, this irregularity prevents the perturbed
bubble from evolving a stable, toroidal structure. In terms of the
overall energy release, the toroidal shape seems to be a favorable
configuration. For the spherical bubble models, as soon as irregular
features form on top of the torus, the energy release increases
dramatically. This boost in the burning rate is much weaker for the
perturbed case.

In our 3D simulations, nuclear burning was suppressed
three seconds after ignition. In all models but 3P50d100, it actually
ceased earlier, since the fuel density in front of the flame fell
below the threshold for burning, $10^7 \, \mathrm{g} \,
\mathrm{cm}^{-3}$. For Model 3P50d100, however, burning was still
active at $t = 3 \, \mathrm{s}$. Therefore, the nuclear energy release
is only a lower limit for this calculation, and the maximum temperature in
the collision (found at $t = 4.81 \, \mathrm{s}$) is an upper bound.

\subsection{Irregular asymmetric and dipolar ignitions}

As in the 2D simulations, irregular one- and two-sided
teardrop-shaped initial flame setups were employed. The evolution of
the two-sided flame ignition case (Model 3T2d200) is shown in
Fig.~\ref{fig:td3d}. Once more, the flame propagates in two opposite
directions. Unlike the 2D simulations, however, the
burning releasing $8.16 \times 10^{50} \, \mathrm{erg}$ is sufficient 
and unbind the star. The shape shown in the right panel of
Fig.~\ref{fig:td3d} marks the final stage of the evolution approaching
homologous expansion. No strong
collision of surface material occurred.

Model 3T1d200, initiated with a one-sided teardrop-shaped flame,
released about two-thirds of the WD's binding energy. As with the
other bubble ignitions, the burning material floated towards the
surface, swept around the core of the WD, and clashed on the opposite
side. The collision parameters are listed in Table~\ref{tab:3d}.

\subsection{Summary of the 3D simulations}
\label{sect:3dconcl}

The diversity of results found in 3D simulations is
larger than that in 2D simulations. The measured
quantities are summarized in Table~\ref{tab:3d}. In
two of the models the WD was even unbound. As plotted in
Fig.~\ref{fig:et1}, all peak collision temperatures in
3D simulations were lower than those in 2D.

A direct comparison between 2D and 3D models is
difficult since models with similar initial flames (in spherical or
teardrop-like shapes) release
significantly more energy when performed in 3D.  The
only way we found to lower the energy release here was by explicitly
perturbing the initial bubble. The two corresponding models, 3P25d100
and 3P50d100, releasing $1.01 \times 10^{50} \, \mathrm{erg}$ and $1.79
\times 10^{50} \, \mathrm{erg}$ of energy in burning, can be compared
to 2D simulations releasing similar amounts of nuclear
energy. The closest examples would be 2B50d200b ($0.97 \times 10^{50}
\, \mathrm{erg}$) and 2B50d100 ($1.93 \times 10^{50} \,
\mathrm{erg}$). These achieved collision temperatures of $2.92 \times
10^9 \, \mathrm{K}$ and $2.00 \times
10^9 \, \mathrm{K}$, respectively, while the collision temperatures
for the 3D models were significantly lower ($1.34 \times
10^9 \, \mathrm{K}$ and $0.724 \times
10^9 \, \mathrm{K}$). 

Two interpretations are possible here.  Either the collision
temperatures are generally lower in 3D due to the
additional degree of freedom decreasing the focusing, or this may only
be the case for models started with strong perturbations from a
spherical bubble since here the toroidal structure supporting focusing
is suppressed. In the first case, the relation between collision
temperature and energy release would be swallower for the 3D data
points in Fig.~\ref{fig:et1} than the relation for the 2D data
points. In the second case, the two 3D data points for the highest
energy releases would fall onto the relation for the 2D sample and the
two low-energy 3D models would diverge from it.

\section{Conditions for detonation}
\label{sect:sdet}

\begin{table}
\begin{center}
\caption{Constraints on detonation initiations.
\label{tab:detconstr}}
\setlength{\extrarowheight}{2pt}
\begin{tabular}{lllll}
\tableline\tableline

\multicolumn{1}{p{5.78em}}{$\rho$ [$10^6 \, \mathrm{g} \, \mathrm{cm}^{-3}$]} &
\multicolumn{1}{p{4.6em}}{$T_\mathrm{c}$ [$10^{9} \, \mathrm{K}$]}&
\multicolumn{1}{p{5em}}{$M$ [g]}&
\multicolumn{1}{p{4.2em}}{$R$ [km]}&
\multicolumn{1}{p{5.0em}}{detonation?}\\
\tableline
10 & 2.6 & $2.5 \times 10^{23}$ & 2 & no \\
10 & 2.7 & $2.5 \times 10^{23}$ & 2 & no \\
10 & 2.8 & $2.5 \times 10^{23}$ & 2 & yes \\
10 & 2.1 & $2.0 \times 10^{25}$ & 8 & no \\
10 & 2.2 & $2.0 \times 10^{25}$ & 8 & yes \\
10 & 1.8 & $1.5 \times 10^{27}$ & 30 & no \\
10 & 1.9 & $1.5 \times 10^{27}$ & 30 & yes \\
\tableline
3 & 2.2 & $2.0 \times 10^{28}$ & 120 & no \\
3 & 2.3 & $2.0 \times 10^{28}$ & 120 & yes \\
\tableline
1 & 2.4 & $3.0 \times 10^{27}$ & 90 & no \\
1 & 3.0 & $3.0 \times 10^{27}$ & 90 & no \\
1 & 3.0 & $3.0 \times 10^{30}$ & 900 & no \\
\tableline
\end{tabular}
\end{center}
\end{table}

\subsection{Constraints on detonation ignitions in degenerate C+O matter}

In order to trigger a detonation a region must burn supersonically and
the size of that region must be larger than some critical
mass \citep{niemeyer1997b,dursi2006a}.  Because that critical mass is
very sensitive to the density and composition, detonation becomes
increasingly difficult at low density and is sensitive to the carbon
mass fraction in the unburned fuel. 

A series of calculations was carried out offline to study the
conditions for detonation using the \textsc{Kepler} 1D hydrodynamics code
\citep{weaver1978}. The procedure was identical to that described in
\citet{niemeyer1997b}. 

A sphere composed of 50\% by mass carbon and 50\% oxygen of prescribed
density, $\rho$, was given a temperature profile characterized by a
central value, $T_\mathrm{c}$, and a linear decline over a specified
range of mass, $M$, defining a radius, $R$, of the sphere. The sphere
was then allowed to runaway inside of a much larger, cooler isothermal
region to see if a successful detonation resulted.

Table~\ref{tab:detconstr} gives the results. If the compression heats
fuel with density above 10$^7$ g cm$^{-3}$ to a temperature over $1.9
\times 10^9$ K on a length scale of 10 km or more (the grid
resolution) detonation will occur.  By $3 \times 10^6$ g cm$^{-3}$,
the necessary temperature has risen to about $2.3 \times 10^9$ K on a
scale of 100 km, and by $\rho = 1 \times 10^6$, \emph{it is impossible
  to detonate the star no matter what temperature is achieved in the
  collision}. The critical mass has become more than a substantial
fraction of the entire star. Even though burning might occur with a
supersonic phase velocity, detonation of carbon does not happen below
10$^6$ g cm$^{-3}$.

\subsection{Comparison with simulations}

\begin{figure}
\includegraphics[width=\linewidth]{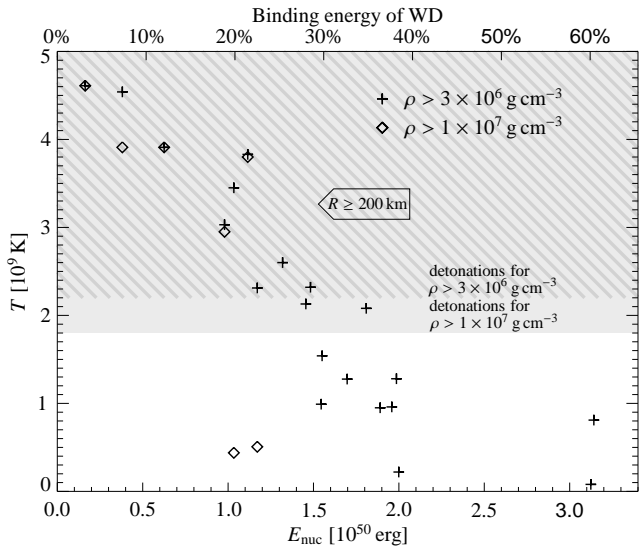}
\caption{\label{fig:et2} Maximum temperature reached in the collision
  region for fuel above the given density thresholds as a function of
  the nuclear energy release prior to reaching the maximum collision
  temperature (according to Tables~\ref{tab:res}, \ref{tab:radvar} and
  \ref{tab:morph}). The shaded and dashed regions correspond to
  conditions 
  where an initiation of a detonation is possible for temperatures
  reached in fuel of densities $\rho > 1 \times 10^7 \, \mathrm{g} \,
  \mathrm{cm}^{-3}$ and $\rho > 3 \times 10^6 \, \mathrm{g} \,
  \mathrm{cm}^{-3}$, respectively.}
\end{figure}

Since it is the combination of temperature and density reached in a
critical mass that decides whether detonation occurs, the maximum
temperatures in the collision region were measured in our 2D
simulations at a density exceeding $1\times 10^7 \,
\mathrm{g}\,\mathrm{cm}^{-3}$ and $3\times 10^6 \,
\mathrm{g}\,\mathrm{cm}^{-3}$, respectively. Values are given in
Tables~\ref{tab:res}, \ref{tab:radvar}, and \ref{tab:morph}, and the
results are plotted in Fig.~\ref{fig:et2}.  Densities in the
compressed material above $10\times 10^7 \, \mathrm{g} \,
\mathrm{cm}^{-3}$ are only found in some rare cases in which the
nuclear energy release prior to collision was lower than $1.2 \times
10^{50} \, \mathrm{erg}$. For all simulations with an energy release
lower than $2 \times 10^{50} \, \mathrm{erg}$, the compression density
of the unburned material in the collision exceeded $\sim$$3\times 10^6
\, \mathrm{g} \, \mathrm{cm}^{-3}$ and the temperature was determined
there.

From these measurements, we find that \emph{a detonation is admissible
  in some of the 2D models}. In models with single-bubble ignitions
displaced more than $200 \, \mathrm{km}$ off-center, sufficient
temperatures are reached at densities exceeding the threshold of $3
\times 10^6 \, \mathrm{g} \, \mathrm{cm}^{-3}$.  The spatial extent of
the compressed region is not critical, since it is, at most, several
grid cells in all cases (typical cell sizes in the collision region
are several km). In models ignited at distances $\ge 400 \,
\mathrm{km}$ off-center, temperatures above $2.5 \times 10^{9} \,
\mathrm{K}$ were reached even at densities above $10^7 \, \mathrm{g}
\, \mathrm{cm}^{-3}$, rendering a detonation virtually certain. A
bubble displacement around $200 \, \mathrm{km}$ marks the bifurcation
value, where secondary parameters, such as resolution, determine the
feasibility of a detonation (cf.\ Table~\ref{tab:res}). It seems
unlikely that the flame ignition takes place at such large radii in
SNe~Ia, but similar results might come from a model with less
efficient lateral burning ignited closer in.

In all 3D simulations, the maximum compression temperatures in the
collision (if one occurs at all) are too low to initiate a
detonation. Moreover, these peak temperatures were found in material
of densities falling short of the detonation threshold for at least
one order of magnitude (cf.\ Table~\ref{tab:3d}). The compressed
region did not reach densities above $10^6 \, \mathrm{g} \,
\mathrm{cm}^{-3}$ near temperature maximum in any of the simulations,
the reasons thereof being the same as those discussed in
Sect.~\ref{sect:3dconcl}. Even the 3D simulations releasing similar
amounts of energy as some 2D simulations favoring a detonation did not
reach the necessary densities. In the two-sided ignition
Model 3T2d200 the the effect of an increased energy release was even
greater than its the 2D analog---it unbound the star.  Thus, all 3D
simulations clearly fail to trigger a detonation.

\section{Conclusions}
\label{sect:concl}

The evolution of thermonuclear supernova models that ignite
asymmetrically has been followed in two and three
dimensions. Parameters of the setup, such as the numerical resolution,
the displacement of the center of flame ignition from the center of
the WD, and the ignition shape were explored in a systematic way in
2D, and the more interesting cases were explored in 3D.

For all 2D simulations, the energy release in the nuclear burning
falls short of unbinding the star. In the cases with one-sided
ignition, the flame floats rapidly to the surface \citep[see
  also,][]{livne2005a}, spreading laterally as it goes due to burning
and instabilities. Since the star is still gravitationally bound, the
emerging ashes sweep around the core of the WD and collide on the
opposite side, in agreement with \citet{plewa2004a}. The collision
strength, and thus the maximum temperature reached in the compressed
fuel correlate inversely with the nuclear energy released on the
flame's way to the surface.  The question of whether the compression
of unburned material in the collision region is adequate to trigger a
detonation was explored in detail, and the necessary criteria were set
out. A detonation can only occur if the fuel temperature exceeds
approximately $1.9 \times 10^9 \, \mathrm{K}$ at a density above
$10^7\, \mathrm{g} \, \mathrm{cm}^{-3}$. At lower densities,
detonation requires higher temperatures and eventually becomes
impossible, for any temperature, for densities less than about $10^6
\, \mathrm{g} \, \mathrm{cm}^{-3}$.

The conditions for initiating a detonation were met in several 2D
calculations in which the flame ignited in a spherical bubble more
than $200 \, \mathrm{km}$ off-center.  Less efficient prescriptions
for the burning might have found similar conditions in simulations
that ignited closer in. For the initial conditions and flame
propagation model assumed, the results of \citet{plewa2004a}, may be
reasonable. However, since the flame model applied there is not based
on a consistent treatment of the flame's interaction with the
turbulent cascade, it is difficult to judge its validity.

\emph{In three dimensions, all simulations fell far short of
  initiating a detonation}. Some even released sufficient energy to
unbind the star. There are several reasons for this
difference. Lacking the artificial symmetry of 2D simulations, the
focusing of the collision in 3D models can be weaker. Indications for
this were found in our simulations. At least two of the 3D simulations
gave significantly lower collision temperatures than predicted by 2D
models that burned similar amounts of fuel (see Fig.~\ref{fig:et1}).
A second, probably dominant effect is that 3D models release more
energy than their 2D analogs. This is mostly due to the additional
degree of freedom in developing flame surface area due to
instabilities. Another contribution to the difference may be the
improved subgrid-scale turbulence model applied in the 3D
simulations. However, this effect is expected to be minor
\citep{schmidt2006c}. An asymmetrically ignited 3D model based on a
different flame implementation \citep{calder2004a} burned
$\sim$$0.075\, M_\odot$ of material. This corresponds to a nuclear
energy release of $\sim$$1.2 \times 10^{50} \, \mathrm{erg}$---a value
that falls in the range spanned by our parameter study. Therefore,
although the Calder et al. model was not followed beyond the breakout
of the ashes from the surface, one expects that that model would also
have failed to trigger a detonation.

While we found no example of a successful detonation in our 3D
simulations, this possibility cannot be completely ruled out since the
exploration of the parameter space was incomplete. An interesting
possibility is that of double-sided ignitions, which
may be possible in a teardrop-shaped ignition overshooting through the
WD's center (or, in a simpler configuration, as two opposed
bubbles). Such a configuration shortens the way the material has to
travel towards the collision spot once ashes break out of the star's
surface on both sides. Therefore the expansion of the star may not be
as advanced as in the collision on the far side of a single-bubble
breakout. In this case, higher temperatures and densities are expected
in the compressed fuel. On the other hand, burning on both sides of
the star releases more energy while the flames propagate towards the
surface. This increases the expansion prior to collision. In our
simulations, the latter effect dominated and the collision was weak in
a 2D model. A similar 3D model became unbound and no detonation was
found in either. Thus, if a multiple surface breakout scenario is to
work at all, no more than two widely separated ignition kernels are
admissible or there will be too much expansion. An open question is
whether a special placement of two or three bubbles spanning a smaller
angle than $180^{\circ}$ with the WD's center might favor the first
effect, increasing the collision strength by shortening the path of the
surface material. However, the bubbles cannot be too close or they
would merge quickly due to burning, without significantly compressing
the unburned material between them, and there cannot be very many, or
they will prematurely unbind the star.

Keeping in mind the uncertainties of the flame model and the
incompleteness of the parameter space explored in 3D
simulations, we conclude, that although a detonation due to the
colliding surface material may, in principle, occur for
certain---possibly artificial---ignition configurations, it cannot
serve as a robust model for SNe~Ia. The simulations presented here
indicate that it may not be realized in nature at all.

For models that remain gravitationally bound, failure to initiate a
detonation will lead to pulsations of the WD star
\citep{nomoto1976a}. This may be a
second chance for triggering a detonation
\citep{arnett1994a,bravo2006a} and this occurrence will be addressed in
a follow-up study.

\acknowledgements{This research used resources of the National Center
  for Computational Sciences at Oak Ridge National Laboratory, which
  is supported by the Office of Science of the U.S. Department of
  Energy under Contract No. DE-AC05-00OR22725. The work was supported
  by the NASA theory program (NNG05GG08G)and the SciDAC Program of the
  DOE (DE-FC02-01ER41176).}


\begin{thebibliography}{45}
\expandafter\ifx\csname natexlab\endcsname\relax\def\natexlab#1{#1}\fi

\bibitem[{{Arnett} \& {Livne}(1994)}]{arnett1994a}
{Arnett}, D., \& {Livne}, E. 1994, \apj, 427, 315

\bibitem[{{Blinnikov} {et~al.}(2006){Blinnikov}, {R{\"o}pke}, {Sorokina},
  {Gieseler}, {Reinecke}, {Travaglio}, {Hillebrandt}, \&
  {Stritzinger}}]{blinnikov2006a}
{Blinnikov}, S.~I., {R{\"o}pke}, F.~K., {Sorokina}, E.~I., {Gieseler}, M.,
  {Reinecke}, M., {Travaglio}, C., {Hillebrandt}, W., \& {Stritzinger}, M.
  2006, \aap, 453, 229

\bibitem[{{Bravo} \& {Garc{\'{\i}}a-Senz}(2006)}]{bravo2006a}
{Bravo}, E., \& {Garc{\'{\i}}a-Senz}, D. 2006, \apjl, 642, L157

\bibitem[{{Calder} {et~al.}(2004){Calder}, {Plewa}, {Vladimirova}, {Lamb}, \&
  {Truran}}]{calder2004a}
{Calder}, A.~C., {Plewa}, T., {Vladimirova}, N., {Lamb}, D.~Q., \& {Truran},
  J.~W. 2004, astro-ph/0405126

\bibitem[{{Colella} \& {Woodward}(1984)}]{colella1984a}
{Colella}, P., \& {Woodward}, P.~R. 1984, J. Comp. Phys., 54, 174

\bibitem[{{Damk{\"o}hler}(1940)}]{damkoehler1940a}
{Damk{\"o}hler}, G. 1940, Z. f. Elektroch., 46, 601

\bibitem[{{Dursi} \& {Timmes}(2006)}]{dursi2006a}
{Dursi}, L.~J., \& {Timmes}, F.~X. 2006, \apj, 641, 1071

\bibitem[{{Filippenko}(1997)}]{filippenko1997a}
{Filippenko}, A.~V. 1997, \araa, 35, 309

\bibitem[{{Fryxell} {et~al.}(1989){Fryxell}, {M{\"u}ller}, \&
  {Arnett}}]{fryxell1989a}
{Fryxell}, B.~A., {M{\"u}ller}, E., \& {Arnett}, W.~D. 1989, Hydro\-dynamics
  and nuclear burning, MPA Green Report 449, Max-Planck-Institut f\"ur
  Astrophysik, Garching

\bibitem[{{Gamezo} {et~al.}(2003){Gamezo}, {Khokhlov}, {Oran}, {Chtchelkanova},
  \& {Rosenberg}}]{gamezo2003a}
{Gamezo}, V.~N., {Khokhlov}, A.~M., {Oran}, E.~S., {Chtchelkanova}, A.~Y., \&
  {Rosenberg}, R.~O. 2003, Science, 299, 77

\bibitem[{{Garc{\'{\i}}a-Senz} \& {Bravo}(2005)}]{garcia2005a}
{Garc{\'{\i}}a-Senz}, D., \& {Bravo}, E. 2005, \aap, 430, 585

\bibitem[{{Garcia-Senz} \& {Woosley}(1995)}]{garcia1995a}
{Garcia-Senz}, D., \& {Woosley}, S.~E. 1995, \apj, 454, 895

\bibitem[{{Hillebrandt} \& {Niemeyer}(2000)}]{hillebrandt2000a}
{Hillebrandt}, W., \& {Niemeyer}, J.~C. 2000, \araa, 38, 191

\bibitem[{{H\"oflich} \& {Stein}(2002)}]{hoeflich2002a}
{H\"oflich}, P., \& {Stein}, J. 2002, \apj, 568, 779

\bibitem[{{Iapichino} {et~al.}(2006){Iapichino}, {Br{\"u}ggen}, {Hillebrandt},
  \& {Niemeyer}}]{iapichino2006a}
{Iapichino}, L., {Br{\"u}ggen}, M., {Hillebrandt}, W., \& {Niemeyer}, J.~C.
  2006, \aap, 450, 655

\bibitem[{{Khokhlov}(1991)}]{khokhlov1991a}
{Khokhlov}, A.~M. 1991, \aap, 245, 114

\bibitem[{{Kozma} {et~al.}(2005){Kozma}, {Fransson}, {Hillebrandt},
  {Travaglio}, {Sollerman}, {Reinecke}, {R{\"o}pke}, \&
  {Spyromilio}}]{kozma2005a}
{Kozma}, C., {Fransson}, C., {Hillebrandt}, W., {Travaglio}, C., {Sollerman},
  J., {Reinecke}, M., {R{\"o}pke}, F.~K., \& {Spyromilio}, J. 2005, \aap, 437,
  983

\bibitem[{{Kuhlen} {et~al.}(2006){Kuhlen}, {Woosley}, \&
  {Glatzmaier}}]{kuhlen2006a}
{Kuhlen}, M., {Woosley}, S.~E., \& {Glatzmaier}, G.~A. 2006, \apj, 640, 407

\bibitem[{{Livne} {et~al.}(2005){Livne}, {Asida}, \&
  {H{\"o}flich}}]{livne2005a}
{Livne}, E., {Asida}, S.~M., \& {H{\"o}flich}, P. 2005, \apj, 632, 443

\bibitem[{{Niemeyer}(1999)}]{niemeyer1999a}
{Niemeyer}, J.~C. 1999, \apj, 523, L57

\bibitem[{{Niemeyer} \& {Hillebrandt}(1995)}]{niemeyer1995b}
{Niemeyer}, J.~C., \& {Hillebrandt}, W. 1995, \apj, 452, 769

\bibitem[{{Niemeyer} {et~al.}(1996){Niemeyer}, {Hillebrandt}, \&
  {Woosley}}]{niemeyer1996a}
{Niemeyer}, J.~C., {Hillebrandt}, W., \& {Woosley}, S.~E. 1996, \apj, 471, 903

\bibitem[{{Niemeyer} \& {Woosley}(1997)}]{niemeyer1997b}
{Niemeyer}, J.~C., \& {Woosley}, S.~E. 1997, \apj, 475, 740

\bibitem[{{Nomoto} {et~al.}(1976)}]{nomoto1976a}
{Nomoto}, K., {Sugimoto}, D., \& {Neo}, S. 1976, \apss, 39, L37

\bibitem[{{Osher} \& {Sethian}(1988)}]{osher1988a}
{Osher}, S., \& {Sethian}, J.~A. 1988, J. Comp. Phys., 79, 12

\bibitem[{{Plewa} {et~al.}(2004){Plewa}, {Calder}, \& {Lamb}}]{plewa2004a}
{Plewa}, T., {Calder}, A.~C., \& {Lamb}, D.~Q. 2004, \apjl, 612, L37

\bibitem[{{Pocheau}(1994)}]{pocheau1994a}
{Pocheau}, A. 1994, \pre, 49, 1109

\bibitem[{{Reinecke} {et~al.}(2002{\natexlab{a}}){Reinecke}, {Hillebrandt}, \&
  {Niemeyer}}]{reinecke2002b}
{Reinecke}, M., {Hillebrandt}, W., \& {Niemeyer}, J.~C. 2002{\natexlab{a}},
  \aap, 386, 936

\bibitem[{{Reinecke} {et~al.}(2002{\natexlab{b}}){Reinecke}, {Hillebrandt}, \&
  {Niemeyer}}]{reinecke2002d}
---. 2002{\natexlab{b}}, \aap, 391, 1167

\bibitem[{{Reinecke} {et~al.}(1999){Reinecke}, {Hillebrandt}, {Niemeyer},
  {Klein}, \& {Gr{\" o}bl}}]{reinecke1999a}
{Reinecke}, M., {Hillebrandt}, W., {Niemeyer}, J.~C., {Klein}, R., \& {Gr{\"
  o}bl}, A. 1999, \aap, 347, 724

\bibitem[{{R{\"o}pke}(2005)}]{roepke2005c}
{R{\"o}pke}, F.~K. 2005, \aap, 432, 969

\bibitem[{{R{\"o}pke} {et~al.}(2006{\natexlab{a}}){R{\"o}pke}, {Gieseler},
  {Reinecke}, {Travaglio}, \& {Hillebrandt}}]{roepke2006b}
{R{\"o}pke}, F.~K., {Gieseler}, M., {Reinecke}, M., {Travaglio}, C., \&
  {Hillebrandt}, W. 2006{\natexlab{a}}, \aap, 453, 203

\bibitem[{{R{\"o}pke} \& {Hillebrandt}(2004)}]{roepke2004c}
{R{\"o}pke}, F.~K., \& {Hillebrandt}, W. 2004, \aap, 420, L1

\bibitem[{{R{\"o}pke} \& {Hillebrandt}(2005{\natexlab{a}})}]{roepke2005b}
---. 2005{\natexlab{a}}, \aap, 431, 635

\bibitem[{{R{\"o}pke} \& {Hillebrandt}(2005{\natexlab{b}})}]{roepke2005a}
---. 2005{\natexlab{b}}, \aap, 429, L29

\bibitem[{{R{\"o}pke} {et~al.}(2006{\natexlab{b}}){R{\"o}pke}, {Hillebrandt},
  {Niemeyer}, \& {Woosley}}]{roepke2006a}
{R{\"o}pke}, F.~K., {Hillebrandt}, W., {Niemeyer}, J.~C., \& {Woosley}, S.~E.
  2006{\natexlab{b}}, \aap, 448, 1

\bibitem[{{Schmidt} \& {Niemeyer}(2006)}]{schmidt2006a}
{Schmidt}, W., \& {Niemeyer}, J.~C. 2006, \aap, 446, 627

\bibitem[{{Schmidt} {et~al.}(2006{\natexlab{a}}){Schmidt}, {Niemeyer}, \&
  {Hillebrandt}}]{schmidt2006b}
{Schmidt}, W., {Niemeyer}, J.~C., \& {Hillebrandt}, W. 2006{\natexlab{a}},
  \aap, 450, 265

\bibitem[{{Schmidt} {et~al.}(2006{\natexlab{b}}){Schmidt}, {Niemeyer},
  {Hillebrandt}, \& {R{\"o}pke}}]{schmidt2006c}
{Schmidt}, W., {Niemeyer}, J.~C., {Hillebrandt}, W., \& {R{\"o}pke}, F.~K.
  2006{\natexlab{b}}, \aap, 450, 283

\bibitem[{{Travaglio} {et~al.}(2004){Travaglio}, {Hillebrandt}, {Reinecke}, \&
  {Thielemann}}]{travaglio2004a}
{Travaglio}, C., {Hillebrandt}, W., {Reinecke}, M., \& {Thielemann}, F.-K.
  2004, \aap, 425, 1029

\bibitem[{{Weaver} {et~al.}(1978){Weaver}, {Zimmerman}, \&
  {Woosley}}]{weaver1978}
{Weaver}, T.~A., {Zimmerman}, G.~B., \& {Woosley}, S.~E. 1978, \apj, 225, 1021

\bibitem[{{Woosley}(1990)}]{woosley1990a}
{Woosley}, S.~E. 1990, in Supernovae, ed. A.~G. {Petschek} (New York:
  Springer-Verlag), 182--212

\bibitem[{Woosley \& Weaver(1994)}]{woosley1994a}
Woosley, S.~E., \& Weaver, T.~A. 1994, in Les Houches Session {LIV}:
  Supernovae, ed. S.~Bludman, R.~Mochovitch, \& J.~Zinn-Justin (Amsterdam:
  North-Holland), 63--154

\bibitem[{{Woosley} {et~al.}(2004){Woosley}, {Wunsch}, \&
  {Kuhlen}}]{woosley2004a}
{Woosley}, S.~E., {Wunsch}, S., \& {Kuhlen}, M. 2004, \apj, 607, 921

\bibitem[{{Wunsch} \& {Woosley}(2004)}]{wunsch2004a}
{Wunsch}, S., \& {Woosley}, S.~E. 2004, \apj, 616, 1102

\bibitem[{{Zingale} {et~al.}(2005){Zingale}, {Woosley}, {Bell}, {Day}, \&
  {Rendleman}}]{zingale2005b}
{Zingale}, M., {Woosley}, S.~E., {Bell}, J.~B., {Day}, M.~S., \& {Rendleman},
  C.~A. 2005, Journal of Physics Conference Series, 16, 405

\end{thebibliography}
\end{document}